\documentclass[referee]{aa} 
\usepackage{graphicx}
\usepackage{txfonts}
\begin{document}
   \title{Pluto's lower atmosphere structure and methane abundance from high-resolution spectroscopy and stellar occultations}
%

   \author{E. Lellouch
          \inst{1}
          \and
          B. Sicardy\inst{1,2}
	  \and
          C. de Bergh\inst{1}
	  \and
	  H.-U. K\"aufl\inst{3}
	 \and
	  S. Kassi \inst{4}
	  \and
	  A. Campargue \inst{4}
         }

   \institute{LESIA, Observatoire de Paris, 5 place Jules Janssen, 92195 Meudon, France\\
              \email{emmanuel.lellouch@obspm.fr}
\and 
Universit\'e Pierre et Marie Curie, 4 place Jussieu, F-75005 Paris, France;  senior member of the Institut Universitaire de
France
         \and
             European Space Observatory, Karl-Schwarzschild-Strasse 2, D-85748 Garching bei M\"unchen, Germany  
	\and
Laboratoire de Spectrom\'etrie Physique, Universit\'e Joseph Fourier, BP-87, F-38402 St-Martin d'H\`eres Cedex, France
             }

   \date{Received January, 9, 2009; revised January 27, 2009, accepted, January 29, 2009}

 
  \abstract
   {Pluto possesses a thin atmosphere, primarily composed of nitrogen, in which the detection of methane
has been reported.}
   {The goal is to constrain essential but so far unknown parameters of Pluto's atmosphere such as the surface pressure, lower atmosphere thermal
stucture, and methane mixing ratio.}
   {We use high-resolution spectroscopic observations of gaseous methane, and a novel analysis of occultation light-curves.}
   {We show that (i) Pluto's surface pressure is currently in the 6.5-24 $\mu$bar range (ii) the methane mixing ratio is 0.5$\pm$0.1~\%, adequate to explain
Pluto's inverted thermal structure and $\sim$100 K upper atmosphere temperature (iii) a troposphere is not required by our data, but if present,
it has a depth of at most 17 km, i.e. less than one pressure scale height; in this case methane is supersaturated in most of it. The atmospheric and bulk surface abundance of methane are strikingly similar, a possible consequence of the presence of a CH$_4$-rich top surface layer. }
   {}

 
\titlerunning{Pluto's lower atmosphere structure and methane abundance} 
   \maketitle
%

\section{Introduction}
Since its detection in the 1980s (Brosch, 1995, Hubbard et al. 1988, Elliot et al. 1989), stellar occultations have revealed remarkable features of Pluto's tenuous ($\mu$bar-like) atmosphere. Pluto's upper atmosphere is isothermal (T$\sim$100 K at altitudes above 1215 km from Pluto's center) and has undergone a pressure expansion by a factor of 2 from 1988 to 2002, probably related to seasonal cycles, followed by a stabilization over 2002-2007 
(Sicardy et al. 2003, Elliot et al. 2003, 2007, E. Young et al. 2008). Below the 1215 km level, occultation lightcurves are characterized by
a sharp drop (``kink") in flux, interpreted as due to either a $\sim$ 10 km-thick thermally inverted layer (stratosphere) or  absorption by a low-altitude haze with significant opacity ($>$0.15 in vertical viewing). 
So far, observations of stellar occultations have not  provided constraints on the atmospheric structure at deeper levels, nor on the 
surface pressure. 

While Pluto's atmosphere is predominantly composed of N$_2$, the detection of methane has been reported from 1.7 $\mu$m spectroscopy
(Young et al. 1997), leading to a rough estimate of the CH$_4$ column density (1.2 cm-am within a factor of 3-4). Even before its detection, 
methane had been recognized to be the key heating agent in Pluto's atmosphere, able to produce a sharp thermal inversion (Yelle and Lunine, 1989, 
Lellouch 1994, Strobel et al. 1996). The large uncertainty in the data of Young et al., however, as well as the unknown  N$_2$ column density,
did not allow one to determine the CH$_4$ / N$_2$ mixing ratio. 

We here report on high-quality spectroscopic observations of gaseous CH$_4$ on Pluto, from which we separately determine the column density and
equivalent temperature of methane. Combining this information with a novel analysis of recent occultation lightcurves, we obtain a precise
measurement of the methane abundance, as well as new constraints on the structure of Pluto's lower atmosphere and the surface pressure.

\section{VLT/CRIRES observations}
Pluto observations were obtained with the cryogenic high-resolution infrared echelle spectrograph (CRIRES, K\"aufl et al. 2004) installed on ESO VLT 
(European Southern Observatory Very Large Telescope) 
UT1 (Antu) 8.2 m telescope. CRIRES
was used in adaptive optics mode (MACAO) and with a 0.4'' spectrometer slit. The instrument consists of four Aladdin III InSb arrays. We focussed 
on the 2$\nu_3$ band of methane, covering the 1642-1650, 1652-1659, 1662-1670 and 1672-1680 nm ranges, at a mean spectral resolution of 60,000, almost five times better than in the Young et al. (1997) observations. Observations were acquired on August 1 (UT = 3.10-4.30) and 16 (UT = 0.55-2.20), 2008, corresponding to mean Pluto (East) longitudes of 299$^o$ and 179$^o$ respectively. (We use the orbital convention of Buie et al. (1997) in which the North Pole is currently facing the Sun). Pluto's topocentric Doppler shift was +20.0 and +24.8 km/s (i.e. $\sim$0.11 and $\sim$0.14 nm) on the two dates respectively, ensuring proper separation of the Pluto methane lines from their telluric counterparts.  On each date, we also observed one telluric standard star (HIP 91347 and HIP 87220, respectively). We emphasize here the August 1 data, which have the highest quality.


\section{Inferences on Pluto's lower atmosphere structure and methane abundance}

   \begin{figure*}
   \centering
   \includegraphics[width=14cm]{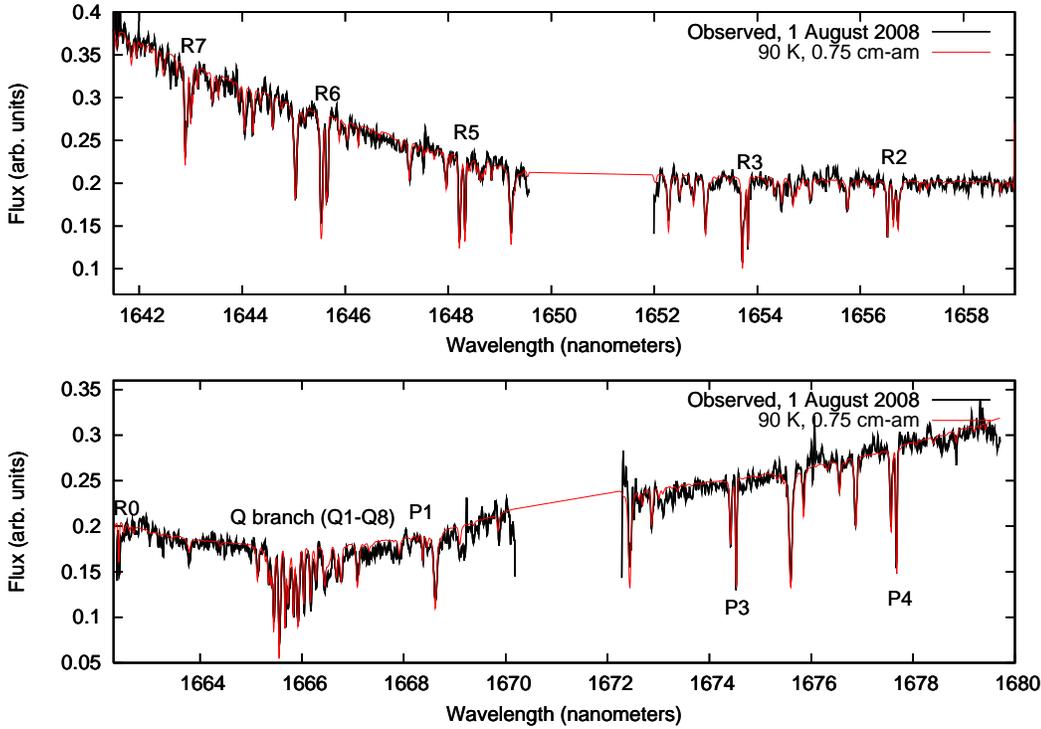}
   \caption{Black: Pluto spectrum observed with VLT/CRIRES. Red: 
Best-fit isothermal model (90 K, 0.75 cm-am CH$_4$), including telluric and solar lines. 
The general continuum shape is due to absorption in the 2$\nu_2$ + $\nu_3$ and 2$\nu_3$ bands of 
solid methane (see Dout\'e et al. 1999)}
              \label{Figallspec}%
    \end{figure*}
The observed spectrum (Fig. ~\ref{Figallspec}) shows the detection of no less than 17 methane lines of the P, Q and R branches of the 2$\nu_3$ band, including high J-level lines (up to R7 and Q8), as well as, more marginally, the presence of a few weaker lines belonging to other band(s) of methane (see below). This spectral richness makes it possible, for the first time, to separate temperature and abundance effects in the Pluto spectra.

\begin{figure}
   \centering
   \includegraphics[width=11cm,angle=0]{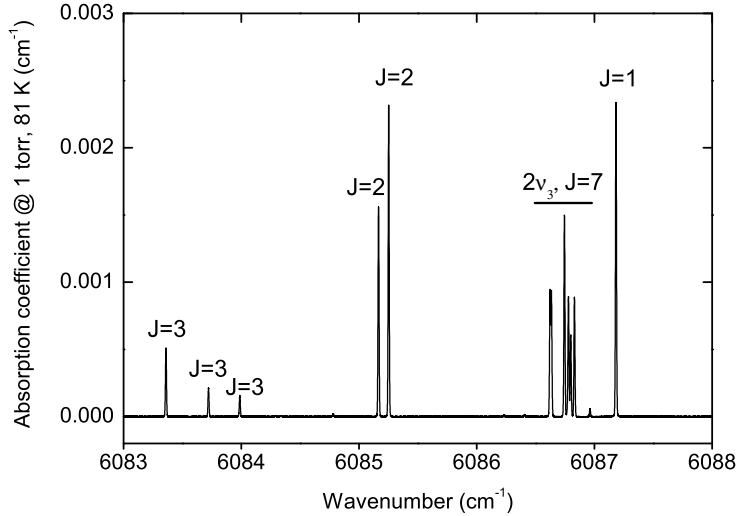}
      \caption{Laboratory spectrum of methane at 81 K in the 6083-6088 cm$^{-1}$ range, demonstrating the existence of strong, low J-level, lines in addition to the R-branch manifolds of the 2$\nu_3$ band. The J-level for these lines is determined by comparison of their intensity at 81 K and at room temperature (see Gao et al. 2009). The J=2 doublet near 6085.2 cm$^{-1}$ is marginally detected in the Pluto spectrum (1643.4 nm, see Fig. ~\ref{Figtemp}).}
         \label{Figspeclabo}
   \end{figure}

Spectra were directly modelled using a telluric transmission spectrum checked against the standard stars observations, a solar spectrum (Fiorenza and Formisano, 2005) and a line-by-line synthetic spectrum of Pluto. The three components were shifted according to their individual Doppler shifts, and then convolved to the instrumental resolution of 60,000, determined by fitting the width of the telluric lines (and corresponding to an effective source size of 0.33''). For modelling the Pluto spectrum, we used a recent CH$_4$ line list (Gao et al. 2009),
based on laboratory measurements (positions and intensities) at 81 K, and including lower energy levels for 845 lines, determined by comparison with the intensities at 296 K collected in the HITRAN database.
Although the temperature of laboratory data is similar to Pluto's, we used only lines for which energy levels were available, in order to avoid dubious extrapolation towards lower temperatures. These data show that, in addition to the J-manifolds of the 2$\nu_3$ band, the spectral range contains other lines of low energy level (e.g. J = 2 near 6085.2 cm$^{-1}$, see Fig. ~\ref{Figspeclabo}), which appear to be marginally detected in the Pluto spectrum (see Fig. 3).

\subsection{Isothermal fits}
\begin{figure*}
   \centering
   \includegraphics[angle=270,width=14cm]{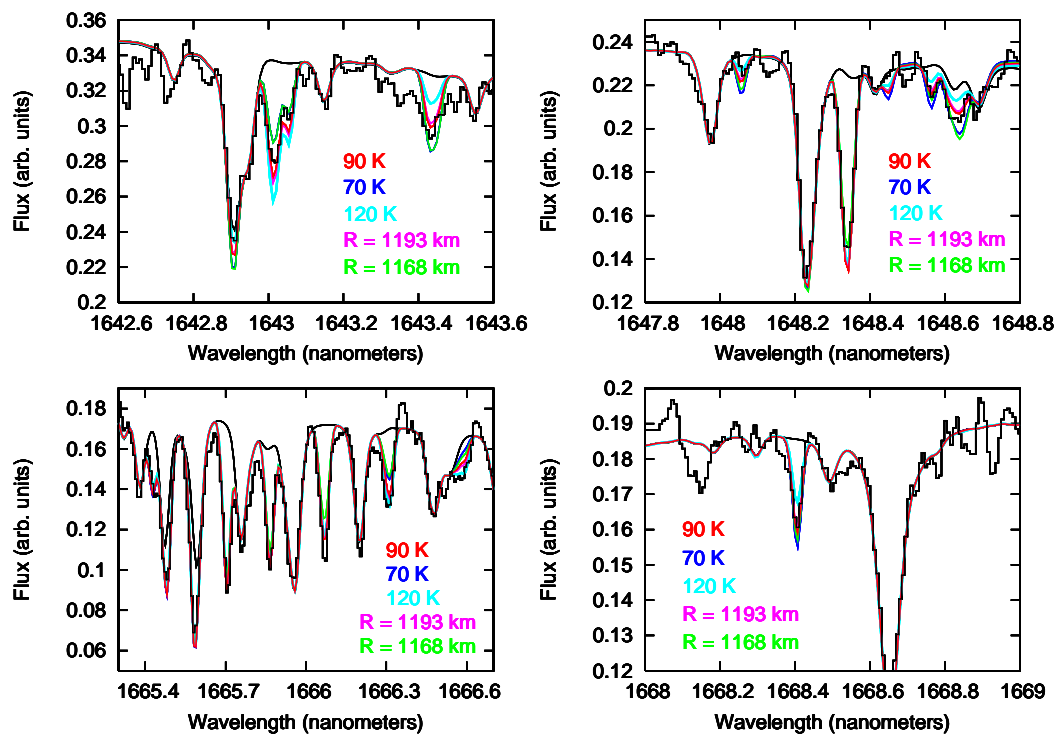}
      \caption{Model fitting of the August 1, 2008 Pluto spectrum (histograms) zoomed on four spectral regions. 
The black curve is a model with no methane on Pluto. The 90 K, 70 K, and 120 K curves indicate isothermal, single-layer, fits, including 0.75 cm-am, 1.3 cm-am and 0.45 cm-am of CH$_4$, respectively. The rotational distribution of lines indicates that a 90 K temperature provides the best fit. The ``R= 1193 km'' model (pink, fit  almost indistinguishable to the 90 K model) corresponds to a 6 K/km stratospheric temperature gradient, a 1193 km radius (7.5 $\mu$bar surface pressure) and a 0.62 \% methane mixing ratio. The ``R=1168 km'' model includes a 6 K/km stratospheric temperature gradient, joining with a wet tropospheric lapse rate of -0.1 K/km below 1188 km (tropopause) and extending down to a 1168 km surface radius (29 $\mu$bar). This 20 km-deep troposphere model, optimized here with CH$_4$ = 0.36 \%, is inconsistent with the methane spectrum; for this thermal profile, the minimum
radius is 1172 km (see Fig. 4). The wavelength scale is in the observer frame. 
These spectral regions are those showing maximum sensitivity to the methane temperature (or equivalently depth of the troposphere), as 
they include low J-level and high J-level lines, but for quantitative analysis, a least-square fit on all lines was performed. 
}
         \label{Figtemp}
   \end{figure*}

We first modelled the data in terms of a single, isothermal methane layer. Because collisional broadening is negligible at the low pressures of Pluto's atmosphere, results at this step are independent of Pluto's pressure-temperature structure. Scattering was ignored, as justified below. The outgoing radiation was integrated over angles, using the classical formulation in which the two-way transmittance is expressed as 2E$_2$ (2$\tau$), where 
$\tau$ is the zenithal optical depth of the atmosphere. 
A least-square analysis of the data was performed in the (temperature (T), column density (a)) space. Fig. ~\ref{Figtemp} shows that the best fit of the Aug. 1, 2008 data is achieved for T = 90 K. Too low (resp. too high) temperatures lead to an underestimate (resp. overestimate) of the high-J lines and an overestimate (resp. underestimate) of the low-J lines.
Based on least-square fitting, we inferred T = 90$^{+25}_{-18}$ K and a = 0.75$^{+0.55}_{-0.30}$ cm-am for the data of August 1, and 
similar numbers (T = 80$^{+25}_{-15}$ K and a = 0.65$^{+0.35}_{-0.30}$ cm-am) for August 16. 

%

\subsection{Combination with inferences from stellar occultations}   
The above inferred methane temperatures, much warmer than Pluto's mean surface temperature ($\sim$50 K, Lellouch et al. 2000) are inconsistent with the existence of a deep, cold and methane-rich troposphere, such as the $\sim$40 km troposphere advocated to match estimates of Pluto's radius from the Pluto-Charon mutual events (Stansberry et al. 1994). To quantify this statement, we combined our spectroscopic data with a new assessment of stellar occultation light-curves. Besides the isothermal part and the ``kink" feature mentioned previously, recent high-quality, occultation curves (Sicardy et al. 2003, Elliot et al. 2003, 2007, E. Young et al. 2008, L. Young et al. 2008) exhibit a number of remarkable characteristics: (i) a low residual flux during
occultation, typically less than 3 \% of the unattenuated stellar flux (ii) the conspicuous absence of caustic spikes in the bottom part of the light-curves (iii) the existence of a central flash, caused by Pluto's limb curvature, in occultations in which the Earth passed near the geometric centre of the shadow.  

To determine the range of Pluto's thermal structures that can account for these features, we performed ray-tracing calculations for a variety of temperature/pressure profiles, expanding upon the work of Stansberry et al. (1994). For this task, we assumed a clear atmosphere. This is justified by (i) the absence of colour variation in the central flash (L. Young et al. 2008) and (ii) the difficulty for hazes to be produced photochemically   
at the required optical depth 
in a tenuous atmosphere like Pluto's (Stansberry et al. 1989).  We thus adopted the ``stratospheric gradient" interpretation of
the light-curves, and explored a broad range of situations, varying the value of this gradient, the level at which the inversion layer
connects to a troposphere (i.e. the tropopause pressure), and the depth and lapse rate of this troposphere (Fig. 4).

\begin{figure}
   \centering
   \includegraphics[width=14cm,angle=0]{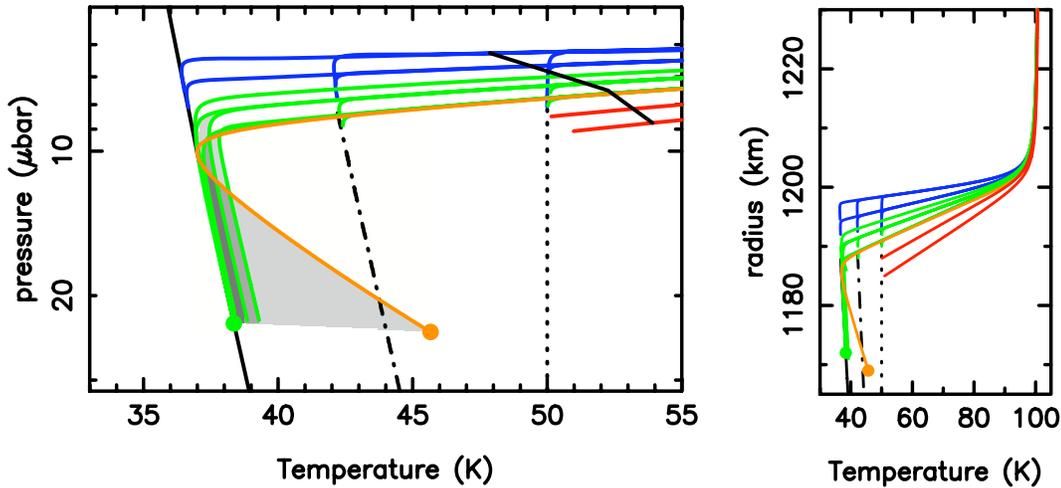}
      \caption{Range of possible thermal profiles (pressure-temperature (left) and radius-temperature (right)) in Pluto's atmosphere.  
From bottom to top, they have stratospheric thermal gradients of 3 and 4 K/km (red profiles), 5 K/km (one orange and one green), 6 K/km
(two green), 7 K/km (green), and 9 and 15 K/km (blue). All profiles are continuous in first and second order temperature derivatives. 
Most of these profiles have no or very limited tropospheres (less than 5 km in depth), in order to match the residual flux observed during stellar occultations and avoid the formation of strong caustics (see Fig. 5). Only profiles in green and orange, with moderate stratospheric temperature gradients (5-7 K/km) and a cold tropopause ($<$ 38 K) can have significant tropospheres. The lapse rate in such tropospheres ranges from -0.1 K/km, corresponding to the N$_2$ wet adiabat (green profiles) to -0.6 K/km (N$_2$ dry adiabat, orange profile). The CRIRES spectra indicate that these wet and dry profiles cannot extend deeper than $p$ $\sim$24 $\mu$bar (1172 and 1169 km, respectively). In the left panel, the solid line on the top right is the locus of minimum atmospheric pressure implied by the observation of a central flash, and the solid line on the left is the vapour pressure equilibrium of N$_2$. The dashed-dotted line is the vapour pressure equilibrium  of  CH$_4$ for a 0.5 \% mixing ratio. The dotted line at 50 K illustrates the maximum possible near-surface gas temperature. The shaded areas represent the range of possible tropospheres. If Pluto has a troposphere, methane must be supersaturated over most of it. 
              }
         \label{Figthermprof}
   \end{figure}

\begin{figure}
   \centering
   \includegraphics[width=14cm,angle=0]{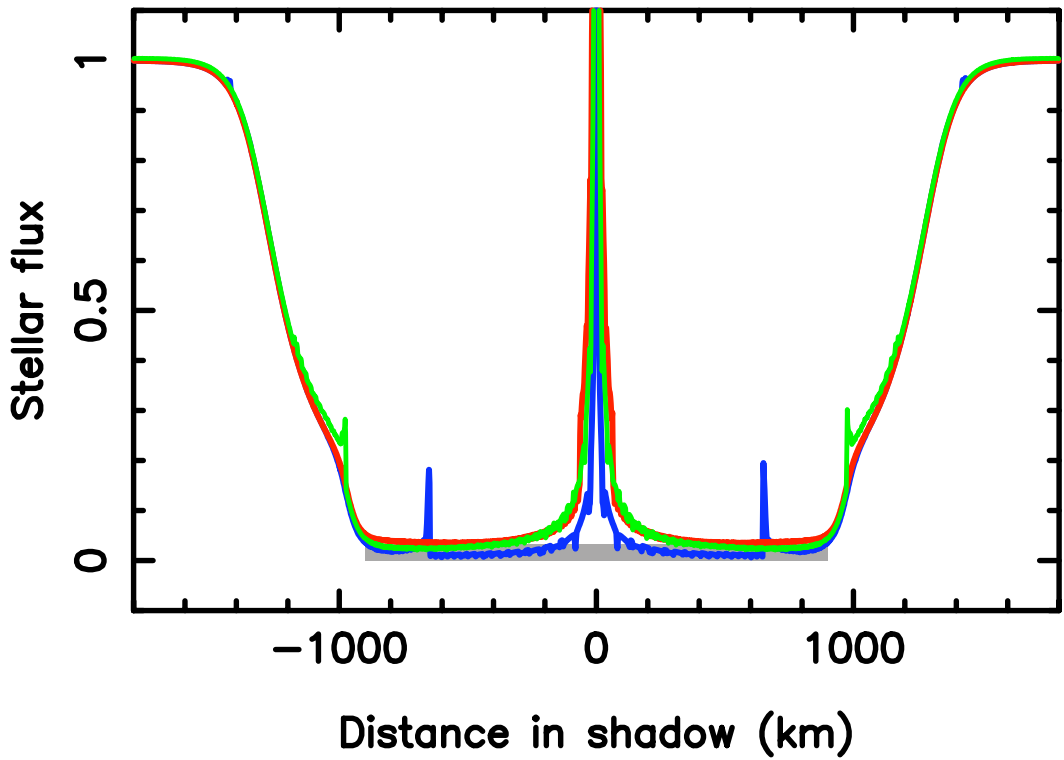}
      \caption{Ray-tracing calculations of occultation light-curves for representative temperature profiles of Fig. 4. The shaded area near the 
bottom of the light-curve  represents the range of residual flux (0.00-0.032) observed in the CFHT August 21, 2002 occultation (Sicardy et al. 2003), with a closest approach to shadow centre of 597 km. (We estimate that the AAT June 12, 2006 occultation light-curve (E. Young et al. 2008) consistently indicates a 0.01-0.03 residual flux). 
Red: light-curve for the thermal profile with 3 K/km stratospheric gradient of Fig. 4, extending to 9 $\mu$bar.  This ``stratosphere-only" model
is consistent with observed light-curves.
Blue: light-curve for the thermal profile with 15 K/km stratospheric gradient, and a 4-km deep troposphere at $\sim$36.5 K. This profile produces
an unacceptable caustics spike, caused by the secondary (``far limb") image 
Green: light-curve for a thermal profile with 6 K/km gradient in the inversion layer, joining the N$_2$ saturation vapour pressure with a $\sim$-0.1 K/km gradient in the troposphere. In this case, modest caustics are still produced, but they appear near the light-curve “kink”.
              }
         \label{Figraytracing}
   \end{figure}

We reached the following conclusions (Fig. ~\ref{Figthermprof} and ~\ref{Figraytracing}): (i) the stratospheric temperature gradient is in the 3-15 K/km range. Gradients smaller than 3 K/km would lead to residual fluxes in excess of 3 \%; gradients larger than 15 K/km produce residual fluxes lower than 1 \%, and are anyway not expected from radiative models (Strobel et al. 1996) (ii) within this range, the existence of the central flash implies a minimum atmospheric pressure of 7.5$\pm$1.2 µbar (iii) the absence of caustic spikes in the region of low residual flux puts stringent constraints on a putative troposphere. In most cases, it restricts such a troposphere to be at most shallow (2-5 km deep, depending on its mean temperature), and the surface pressure to be less than $\sim$10 $\mu$bar. An exception is the family of thermal profiles with intermediate (5-7 K/km) stratospheric temperature gradients and a cold ($<$ 38 K) tropopause, which appear consistent with occultation curves for any tropospheric depth. In fact, such profiles (green curves in Fig. 4 and 5) lead to modest caustic spikes in the region of the ``kink", i.e. where spikes are observed in actual observations, for which they can be mistaken. 

The allowed thermal profiles were finally tested against the methane spectrum. We assumed uniform atmospheric mixing, a plausible case given that (i) the source of methane is at the surface (ii) its equivalent temperature implies that a large fraction of methane is in the upper atmosphere, and performed
a least-square analysis of the data in the (surface radius, CH$_4$ mixing ratio) domain. Not surprisingly in view of the isothermal fits,
thermal profiles having no (or a mini-) troposphere are all consistent with the methane spectrum.
For example, for a stratospheric temperature gradient of 6 K/km,  a surface radius of 1193 km (surface pressure = 7.5 $\mu$bar, i.e. the minimum
required by the occultations) 
provides an adequate fit of the August 1 data for a CH$_4$ mixing ratio of 0.62 \% . In contrast, profiles including too deep a troposphere can be rejected as giving too much weight to cold methane and leading to a line distribution inconsistent with the data. Based on such fits, the maximum tropospheric depth is found to be 17 km (i.e. 0.85 pressure scale heights) and the maximum surface pressure is 24 $\mu$bar. Taking all constraints together, Pluto's surface pressure in 2008 is in the range 6.5-24 $\mu$bar. The range of methane column densities is 0.65-1.3 cm-am. Deeper (i.e. colder) models require larger methane columns than shallower models, but since they also have a higher surface pressure, the methane mixing ratio is accurately determined to be 0.51$\pm$0.11 \%. 
Constraints from the August 16 data are somewhat looser (a maximum surface pressure and troposphere depth of 32 $\mu$bar and 23 km, respectively). 
The minimum Pluto radius implied by the data is 1169-1172 km (Fig. 4). This value holds for the nominal astrometric solutions for 
stellar occultations, typically uncertain by $\sim$10 km. Given this uncertainty, our lower limit on the radius is consistent with most inferences 
from the mutual events (nominally 1151-1178 km, see Tholen et al. 1997). The troposphere depth is free from this uncertainty,
and therefore better constrained than Pluto's radius. 

\section{Discussion}
\subsection{Methane mixing ratio and possible supersaturation}
Through absorption of solar input in the near-IR and radiation at 7.7 $\mu$m, methane is the key heating/cooling agent in Pluto's atmosphere, and in particular
must be responsible for its thermal inversion. Detailed calculations (Strobel et al. 1996) show that, even in the presence of CO cooling and for an assumed 3 $\mu$bar ``surface'' (i.e. base of the inversion layer) pressure, a 0.3 \% methane mixing ratio produces a 7 K/km ``surface'' gradient and a temperature increase of $\sim$36 K in the first 10 km. Although such calculations will need to be redone in the light of our results, a 0.5 \% methane mixing ratio is clearly adequate to explain the $\sim$6 K/km gradient indicated by the occultation data, further justifying our assumption to neglect haze opacity.

The presence of methane in Pluto's stratosphere implies that it is not severely depleted by atmospheric condensation. Yet, a remarkable result is that for models including a troposphere, methane appears to be significantly supersaturated (Fig. ~\ref{Figthermprof}), by as much as a factor $\sim$ 30 for a  $\sim$38 K tropopause. Given that Pluto's troposphere is at most shallow (less than 1 pressure scale height), this plausibly results from convective overshoot associated with dynamical activity, combined with a paucity of condensation nuclei in a clear atmosphere.  

\subsection{The origin of the elevated methane abundance}
In agreement with Young et al. (1997), the CH$_4$ / N$_2$ mixing ratio we derive is orders of magnitude larger than the ratio of their vapour pressures at any given temperature, and the discrepancy is even worse if one considers that methane is a minor component on Pluto's surface. Two scenarios
(Spencer et al. 1997,  Trafton et al. 1997) have been described to explain this elevated methane abundance (i) the formation, through surface-atmosphere exchanges, of a thin methane-rich surface layer (the so-called ``detailed balancing'' layer),  which inhibits the sublimation of the underlying, 
dominantly N$_2$, frost, and leads to an atmosphere with the same composition as this frost (ii) the existence of geographically separated patches of pure methane, warmer than nitrogen-rich regions, and which under sublimation boost the atmospheric methane content. Interestingly, detailed analyses of 1.4-2.5 $\mu$m and 
1-4 $\mu$m mid-resolution spectra give observational credit to both situations. It is noteworthy that our 0.5 \% atmospheric abundance is identical to the CH$_4$ / N$_2$ ratio in the N$_2$ - CH$_4$ - CO subsurface layer of Dout\'e et al. (1999), consistent with the detailed balancing model, and agrees also with the solid methane concentration inferred by Olkin et al. (2007) (0.36 \%). In this framework, a typical 15 $\mu$bar surface pressure could be explained if the N$_2$ - CH$_4$ - CO subsurface layer is at 40.5 K (consistent with the N$_2$ ice temperature measurements of Tryka et al., 1994) and overlaid by a 80 \% CH$_4$ - 20 \% N$_2$ surface layer. On the other hand, and in favour of the alternate scenario, thermal IR lightcurves (Lellouch et al. 2000) as well as sublimation models for CH$_4$ (Stansberry et al. 1996) indicate that extended pure CH$_4$ patches may reach dayside temperatures well in excess of 50 K; this is more than sufficient to explain the $\sim$0.075 $\mu$bar CH$_4$ partial pressure indicated by our data.
 
Discriminating between the two cases may rely on the time evolution of the N$_2$ pressure and CH$_4$ mixing ratio. In particular, the decrease of atmospheric CH$_4$ with increasing heliocentric distance is expected to lead to a drop of the CH$_4$ abundance in the detailed balancing layer, which may delay the decrease of the N$_2$ pressure (Trafton et al. 1998). Assuming T = 100 K, Young et al. (1997)
 reported a 0.33-4.35 cm-am methane abundance in 1992. Although their error bars are very large, their best fit value (1.2 cm-am) is larger than ours (0.65 cm-am for this temperature). Combined with the factor of $\sim$2 pressure increase between 1988 and 2002, this suggests that the methane mixing ratio is currently declining. The ALICE and Rex instruments on {\em New Horizons} will measure Pluto's surface pressure and methane abundance in 2015. Along with the data presented in this paper, this will provide new keys on the seasonal evolution of Pluto's atmosphere and the surface-atmosphere interactions.   
%
%


\begin{acknowledgements}
This work is based on observations performed at the European Southern Observatory (ESO), proposal 381.C-0247. We thank
Darrell Strobel for constructive reviewing.      
\end{acknowledgements}

\end{document}